\documentclass[preprint,showkeys,aps,prb]{revtex4}
\usepackage[dvips]{graphicx}
\begin{document}

FIGURE LIST :\\
1. Black particles in two-ball collision forward and backward\\
2. Three Geometries for generating shockwaves\\
3. 8192-particle twofold shockwave\\
4+5. Runge-Kutta Shockwave reversal---$>$Rarefaction\\
6+7. Shock forward and Shock backward important particles 2048\\
8+9 Rarefaction forward and backward important particles 2048\\
10. Thermodynamics of the 8192 rarefaction wave  

\title{                                                                                                                           
Time-Symmetry Breaking in Hamiltonian Mechanics. II. \\
A Memoir for Berni Julian Alder [1925-2020]                          
}
\author{
William Graham Hoover with Carol Griswold Hoover \\
Ruby Valley Research Institute                   \\
601 Highway Contract 60                          \\
Ruby Valley, Nevada 89833                        \\
}

\date{\today}


\vspace{0.1cm}

\begin{abstract}
This memoir honors the late Berni Julian Alder, who inspired both of us with his pioneering development of molecular dynamics. Berni's work with Tom Wainwright, described in the 1959 Scientific American[1], brought Bill to interview at Livermore in 1962.  Hired by Berni, Bill enjoyed over 40 years' research at the Laboratory. Berni, along with Edward Teller, founded UC's Department of Applied Science in 1963. Their motivation was to attract bright students to use the laboratory's unparalleled research facilities. In 1972 Carol was offered a joint LLNL employee-DAS student appointment at Livermore. Bill, thanks to Berni's efforts, was already a Professor there. Berni's influence was directly responsible for our physics collaboration and our marriage in 1989. The present work is devoted to two early interests of Berni's, irreversibility and shockwaves.  Berni and Tom studied the irreversibility of Boltzmann's ``H function'' in the early 1950s[2]. Berni called shockwaves the ``most irreversible'' of hydrodynamic processes[3].  Just this past summer, in simulating shockwaves with time-reversible classical mechanics, we found that reversed Runge-Kutta shockwave simulations yielded nonsteady rarefaction waves, not shocks.  Intrigued by this unexpected result we studied the exponential Lyapunov 
instabilities in both wave types. Besides the Runge-Kutta and Leapfrog algorithms, we developed a precisely-reversible manybody algorithm based on trajectory storing, just changing the velocities' signs to generate the reversed trajectories. Both shocks and rarefactions were precisely reversed. Separate simulations, forward and reversed, provide interesting examples of the Lyapunov-unstable symmetry-breaking models supporting the Second Law of Thermodynamics.  We describe promising research directions suggested by this work.
\end{abstract}

\maketitle

\section{Introduction}

Bill began to work with Berni in the fall of 1962.  Over the next six years they published six joint works\cite{b4,b5,b6,b7,b8,b9},
including one each with three coauthors: Francis Ree, Tom Wainwright, and Dave Young. All six works were motivated by Berni's
longstanding interest in understanding melting transitions for disks and spheres. The titles give an idea of their joint research:
``Cooperative Motion of Hard Disks Leading to Melting''\cite{b4}; ``Dependence of Lattice Gas Properties on Mesh Size''\cite{b5};
``Cell Theories for Hard Particles''\cite{b6}; ``The Pressure, Collision Rate, and Their Number Dependence for Hard Disks''\cite{b7};
 ``High-Density Equation of State and Entropy for Hard Disks and Spheres''\cite{b8}; and last of all a longer review of their work,
``Numerical Statistical Mechanics'', pages 79-113 in  {\it Physics of Simple Liquids}\cite{b9}, edited by three of their friends
and colleagues: Neville Temperley, John Rowlinson, and George Rushbrooke. These six papers can be found in the chronological
publications list on our website, hooverwilliam.info, under ``[ The 1960s ]''.

Besides introducing us to his worldwide colleagues Berni passed on cogent research advice: understanding is the goal; words
and pictures are vital to understanding; equations, not so much; clarity of presentation is essential; of the three routes to
understanding, formalism, experiment, and computation, at least two of these must be included and compared to make a publication
``useful''.

Our goal in the present work is to shed more light on the connection of time-reversible atomistic dynamics to the irreversible
Second Law of Thermodynamics.  It is an extension of work with a similar title published in 2013\cite{b10}.  Back then, we expressed
our motivation:
\begin{quote}
``The goal we pursue here is improved microscopic understanding of the thermodynamic irreversibility described by the
Second Law of Thermodynamics.''
\end{quote}
In Section II we sketch three approaches to the irreversibility question: [ 1 ] the H Theorem, [ 2 ] fractal distributions from
thermostatted systems, and [ 3 ] time-symmetry breaking through Lyapunov instability. Section III describes the example motivating
the present work, a one-dimensional strong shockwave, simulated with classical manybody molecular dynamics.  The shockwave study led
automatically to an investigation of rarefaction waves. Sections IV and V detail the Lyapunov instabilities of both processes,
shock and rarefaction, in both time directions, ``forward'' and ``backward''. In both cases we develop and apply a novel 
precisely-reversible integration algorithm. Section VI describes the smooth-particle technique for connecting the atomistic and
continuum descriptions of flow problems, applied there to the measurement of longitudinal and transverse temperatures. A summary follows,
in Section VII.

\section{Three Explanations of Dynamical Irreversibility}

In 1956 Berni and Tom described several problems in their Brussels presentation ``Molecular Dynamics by Electronic Computers''\cite{b2}.
Their evaluation of Boltzmann's H Function, the 19th-century explanation of irreversibility, showed that low-density hard-sphere
molecular dynamics and Boltzmann's equation agreed quite well. In 1987 a second explanation of irreversibility from time-reversible
dynamics\cite{b11} was offered as a consequence of Shuichi Nos\'e's equilibrium thermostat ideas\cite{b12,b13} applied to
nonequilibrium problems, following the progress of one- or two-dimensional particles through arrays of scatterers. The time-averaged
temperature was controlled in the one-dimensional case\cite{b14} and the instantaneous temperature was fixed in the two-dimensional
case\cite{b15}. Both these problems supported a new explanation of irreversibility. Both generated fractal phase-space distributions
with fractional dimensionalities less than that of the phase space. The rarity of nonequilibrium states, coupled with the exponential
instability of the reversed fractal repellor motion, provided an explanation more general than Boltzmann's.  Rather than dilute gases
the fractal description applied to a wide variety of liquid and solid problems\cite{b11}.

In 2013 we made a third effort to understand irreversibility for manybody Newtonian systems through a novel measure of Lyapunov
instability\cite{b10}. This pervasive instability can be followed by tracking the rate at which two nearby trajectories, the
``reference'' and the ``satellite'', tend to separate, with the distance, but not the direction,  between the two trajectories held
fixed. The direction of the reference-to-satellite vector joining the two manybody trajectories determines which particles contribute
most to the instability. {\bf Figure 1} shows a striking difference between forward and backward analyses of an inelastic collision
of two 400-particle balls\cite{b10}. The simulation is purely classical and precisely time-reversible.  Forward in time the satellite
particles most sensitive to instability (black in {\bf Figure 1}) are those on the leading edges, those first to take notice of
collision. When precisely the same trajectory is analyzed backward, with the 800-particle ball spontaneously (and completely
unphysically) separating into its two parts, the ``important particles'' are very different.  Backward in time such particles are
mostly in the high-strainrate necking region where new surfaces are being created. The forward collision is physically reasonable and
can be simulated easily with a variety of integrators and algorithms, all of them leading to similar results. The reversed process,
in which a single ball spontaneously separates into parts, is a different story, ``irreversible''. It cannot be simulated directly.
Instead it can only be studied by a brute-force numerical reversal of the forward-in-time collision.

\section{Shockwaves--The ``Most Irreversible'' Processes\cite{b3}}

A comprehensive 1980 study\cite{b16} examined the two shockwaves, with velocities $\pm u_s$, that result when a periodic liquid
manybody system is suddenly compressed by two periodic images of itself. The left image advances rightward  at the ``piston velocity''
$+u_p<u_s$. The right image leftward, at $-u_p$, propelling the faster shock with velocity $-u_s$.  In the space of about two atomic
diameters the argon liquid being modelled increases in pressure to 400 kilobars and in temperature to about ten thousand kelvins. The
density increases approximately twofold.

Here we consider an alternative mechanism for shock generation, and in two space dimensions rather than three.  See the middle
illustration in {\bf Figure 2}. We launch a stress-free cold solid against a fixed barrier at speed $u=u_p$.  When complete, this
process converts the initial macroscopic kinetic energy, $(Nu^2/2)$, into the internal energy of the resulting hot shocked fluid,
$Ne$. We model the initial cold state with an $N$-particle triangular lattice, periodic in $y$.  Each particle pair interacts with
the short-ranged repulsive pair potential, arbitrarily normalized to unity:
$$
\phi(r<1) = (10/\pi)(1-r)^3 \ ; \ \phi(r>1) = 0 \rightarrow \int_0^\infty 2\pi r \phi(r)dr \equiv 1 \ .
$$
In the present shockwave work $N$ is either $8192 = 32 \times 256$ or $2048 = 16 \times 128$ so that the aspect ratio $(L_x/L_y)$, with
close-packed columns of particles parallel to the $y$ axis, is initially $8\sqrt{(3/4)} = 6.9282$. The shock propagation direction
is parallel to the  $x$ axis.

The initial velocity, 0.97, is selected to shock-compress the cold solid twofold, to a hot fluid state. To break the lattice symmetry
we begin with additional thermal velocities corresponding to an otherwise negligible temperature of 0.0001. {\bf Figure 3} shows the
coexistence of the hot shocked material with the cold stress-free triangular-lattice as modelled with 8192 particles. The number
density $\rho$ increases from $\sqrt{(4/3)}$ to $2\sqrt{(4/3)}$ and the internal energy change is consistent with the Hugoniot relation
for twofold compression from the stress-free zero-energy cold state to a hot shocked state with temperature $T_H = 0.115$ :
$$
e_H - e_C \equiv (1/2)(P_H+P_C)(v_C-v_H) \ {\rm [ \ Hugoniot \ Relation \ ] }
$$
$$
{\rm with} \ e_C = 0 {\rm \ and \ } P_C = 0 {\rm \ and \ }  v_H = (v_C/2) \longrightarrow e_H = (1/2)P_H(v_C/2) = (1/2)P_Hv_H \ .
$$
$$
{\rm so \ that \ } e_H = P_H(0.433013/2) = 0.47045 = (0.97^2/2) \rightarrow P_H = 2.173 .
$$
To derive the Hugoniot relation imagine the cold zero-energy zero-pressure crystal moving rightward at speed (0.97/2)
and stagnating to match the velocity of a leftmoving wall at velocity $(-0.97/2)$. In this thought experiment the kinetic
energy of the resulting leftmoving hot fluid is identical to that of the initial cold rightmoving solid, $(1/2)(0.97/2)^2$
per particle. Evidently the resulting internal energy $e_H$ (the energy exclusive of the macroscopic motion) is identical
to the per-particle work done by the crystal in the compression process, $(P_Hv_H/2) = (0.97^2/2)$.

Just as in earlier work\cite{b10} simulations show that the structures of such strong shockwaves are steady and accurately
one-dimensional, with a shockwidth on the order of two particle diameters. In the shock-based coordinate system ( fixed on
the stationary shock, as shown in the top view of {\bf Figure 2} ) cold crystal enters from the left, with $u = u_s = 2u_p$,
and exits at the right with $u = u_s-u_p = u_p = (u_s/2) = 0.97$. A time-reversal of this nonequilibrium shock process is
easily implemented in a Runge-Kutta simulation by changing the sign of the timestep, $dt = 0.01 \rightarrow dt = -0.01$,
or changing the signs of all the velocities in the problem.

{\bf Figures 4 and 5} illustrate the surprising result of this straightforward ``reversal''. It motivated the present work.  Rather
than seeing the shock travel backward unchanged, at least for a reasonable time, instead we found that a rarefaction wave soon
appears. Such a wave is typically generated by the nearly isentropic expansion of a compressed fluid and is discussed in
standard fluid mechanics texts\cite{b17,b18} for simple fluid models.  An accurate Leapfrog integrator, likewise conserving
energy throughout the run to an accuracy of seven digits, produces a similar, likewise surprising, rarefaction. The ``reversed
motion'' generated with either Runge-Kutta integration or Leapfrog is actually anything but!  Notice the holes developing in
the reversed solution. To investigate the mechanism for this convincing failure of algorithmic reversibility we turned to an
analysis of the Lyapunov instability of the process. We expected to see an analog of the symmetry breaking found for two
colliding crystallites as shown in {\bf Figure 1}. We will shortly discuss this investigation, in the next Section, IV. First
we remind the reader how Lyapunov instability is characterized in numerical simulations{\cite{b19,b20,b21}.

\subsection{Lyapunov Instability with a Satellite Simulation}

The largest Lyapunov exponent identifies that part of a system in which the mechanics is least stable, with the highest
growth rate of perturbations.  It is evaluated in practice by following the progress of two neighboring trajectories,
the ``reference'' and the ``satellite'', rescaling their separation at the end of each timestep. The magnitude of this
offset--here we use 0.0001--can be measured in coordinate $q$, momentum $p$, or $(q,p)$ phase space. To carry out a
precisely-reversed simulation one could use either Levesque and Verlet's bit-reversible algorithm\cite{b22} or our
more-nearly-accurate implementation of one of Milne's fourth-order algorithms\cite{b10}.  Both these approaches express
the particle coordinates as (large) integers. Typical force contributions, $\ddot xdt^2$ or $\ddot ydt^2$, become
considerably smaller integers, but are still large relative to unity. Consistent floating-point computations of the force
contributions, truncated to integers, then provide integer coordinate increments which are identical, apart from sign, in
a pair of precisely-reversed motions.

\subsection{A Simpler Time-Reversed Algorithm}

For enhanced accuracy and simplicity we choose here a simpler time-reversible method of simulation, first storing an
accurate Runge-Kutta reference trajectory for thousands of timesteps and then separately computing {\it two} nearby satellite
trajectories, one forward and one reversed.  The offset lengths of both satellite trajectories from the reference are returned
from $|\delta(t)|$ to a fixed length $\delta_0$ at the completion of each timestep, giving the instantaneous value of the largest
Lyapunov exponent,  $\lambda_1(t) \equiv \ln(|\delta(t)|/\delta_0)/dt$, for small $dt$, $\pm 0.01$ in our simulations.
All three trajectories, the reference and two satellites, are generated with the same Runge-Kutta integrator. A novel vital
detail is that the positions of the satellite and reference trajectories often straddle a periodic boundary (in the $y$ direction
when the wave propagation direction is parallel to the $x$ axis). To avoid discontinuous jumps in the vector separating the two
solutions it is necessary to detect and correct satellite coordinates which straddle the boundary, adding or subtracting
$L_y$ as the case may be, resulting in a continuously varying offset vector $\delta(t)$.  

An interesting consequence of the Lyapunov analysis is that the (largest) Lyapunov exponent is uniformly positive in both
time directions. Its numerical value is mostly in the range from 1 to 2 throughout both shockwave and rarefaction
wave simulations. Insight into the Lyapunov instability of the motion comes from identifying which particles
contribute most to the offset vector.  In a pioneering effort Stoddard and Ford\cite{b19} calculated the largest
Lyapunov exponent of a Lennard-Jones fluid in 1967, maintaining the offset in coordinate space.

In 1998, with Kevin Boercker and Harald Posch \cite{b23}, Bill simulated a nonequilibrium field-driven manybody particle flow
and followed the largest local Lyapunov exponent, separately and instantaneously, in coordinate space and momentum space.  The
two identifications of the exponent's ``important particles'' (those with above-average separations, $\delta_x^2 + \delta_y^2$ or
$\delta_{p_x}^2+\delta_{p_y}^2$), were very similar. Nearly all important particles in coordinate space were also important in
momentum space, and {\it vice versa}. One could quantify a particle's contributions to Lyapunov instability in at least three ways,
in terms of
$$
\delta_x^2+\delta_y^2 \ {\rm or} \ \delta_{p_x}^2+\delta_{p_y}^2 \ {\rm or} \
\delta_x^2+\delta_y^2 + \delta_{p_x}^2+\delta_{p_y}^2 \ .
$$
Though different in principle\cite{b24}, all three measures are in practice very similar in the particles they emphasize\cite{b23}.
{\bf Figures 4 and 5} display the result of an important-particle Lyapunov analysis in coordinate space using the straightforward Runge-Kutta
integrator, forward for 6000 timesteps and backward for another 6000, with $dt = \pm0.01$.  Here and in {\bf Figures 6-9} we
use 2048 particles rather than 8192 in order better to visualize details on an individual particle scale.  {\bf Figures 4 and 5}  make the
point quite convincingly that shockwaves are irreversible, even with very accurate integrators.  Let us clarify the meaning of this
observation by {\it storing} the (forward) evolution of the shockwave trajectory and then analyzing it for Lyapunov instability in both
time directions. 

\section{Precisely-Reversible Shock Wave Analyses}

Here {\bf Figures 6 and 7} compare 2048-particle Lyapunov analyses forward and backward for the precisely-reversible (as the coordinates and
momenta are all stored) simulations of that ``most irreversible'' shock process, the process shown in {\bf Figure 3} for 8192 particles.
The configurationally important particles have been colored brown in {\bf Figures 4-9}.  Notice that only in the reversed direction
is the shockwave itself the maximally unstable portion of the system.  Exactly the same configurations, when analyzed forward in
time rather than backward, show that the shockwave is relatively stable (as opposed to unstable) {\it at} the shockfront.  Maximal
instabilities instead occur here and there throughout the hot fluid, in relatively small transient clumps when the propagation is
analyzed forward in time.  Similar clump formation was found in the
field-driven motion analyzed in Reference 23. The difference in the location of ``important particles'' (backward in time, found
 at the shock, but forward in time, located in distant clumps) is a significant positive indication that Lyapunov analyses of
Newtonian mechanics can provide a detailed understanding of the Second Law of Thermodynamics through the measurement of local
instabilities.  By including information local in space and time from past history the Lyapunov offset vectors,
$\{ \ \lambda_1(t\pm dt) \longleftrightarrow \delta_1(t) \ \}$  quantify the simultaneous relative instabilities of microscopic motions.
The difference found here between the forward and backward stability analyses of shocks is qualitative, not just quantitative, in the
shockwave problem. We will come back to this analysis in our Summary section.

\section{Precisely-Reversible Rarefaction Wave Analyses}

In an effort to learn more here, we next generated, analyzed, and studied the evolution of instability in a rarefaction wave. Apparently
the lower-density boundary condition in the reversed version of {\bf Figure 5} provides an unnecessary perturbation of such a wave. To initiate
a simpler pure-rarefaction simulation we first carry out an equilibrium Nos\'e-Hoover\cite{b25} isothermal high-density simulation (2048 particles
with $\rho = 2\sqrt{(4/3)}$ and $T = 0.115$). The resulting equilibrated hot-fluid sample should allow us to start up a rarefaction
simulation in a density-temperature state similar to that reached by the shockwave compression in the forward versions of {\bf Figures
4 and 6}. Rather than using periodic boundaries in both the $x$ and $y$ directions, as is usual in equilibrium situations, here we impose
quartic boundary potentials, $dx^4/4$ at the left and right. These two smooth boundaries repel those particles venturing a distance $dx$
beyond the limits $x = \pm (L_x/2)$. After equilibration, a rarefaction wave should result when we release one of the $x$ boundaries.
We choose to release the righthand boundary.

{\bf Figures 8 and 9} compare the forward and backward instability analyses of the resulting rarefaction wave. To make the details clear we again
use only 2048 particles. The resulting wave was constructed with a three-step process, first simulating 20000 equilibration timesteps at
the high-temperature high-pressure thermodynamic state reached earlier by shock compression. Next, the righthand boundary was released and
the resulting expansion followed for 4000 Runge-Kutta timesteps, a time of 40. Finally, the velocities were reversed for a time of 40,
returning to a close approximation of the initial high-temperature high-pressure state. This preliminary investigation surprised us yet again.
Expansion (forming a rarefaction wave), followed by time reversal, showed no tendency toward shock formation. Instead the reversed flow
closely approximated the rarefaction configurations. To analyze the motion precisely after equilibration, we followed and stored the 4000
$\{ x,y,p_x,p_y \}$ rarefaction states, analyzing them in both directions so as to see the local ``important particles''. {\bf Figures 8 and 9}
shows the important particles found in both time directions for the rarefaction wave. Here the unstable portions of both the forward and
the backward rarefaction flows are all distributed in the hotter denser part of the wave. It is interesting, and was surprising to us,
to see that reversing a rarefaction wave showed no tendency toward shockwave formation.

\section{Continuum Field Variables from $(q,p)$ Particle Information}

{\bf Figure 10} displays thermodynamic data from the stored forward = backward trajectory of {\bf Figures 8 and 9}. The velocities stored for the
latter figure show no essential difference between the longitudinal and transverse temperatures, indicating that the rarefaction wave
is indeed nearly isentropic. Such a wave provides the chance to measure the isentropic equation of state over a range of density and
temperature.  Let us do so now. We calculate ``smoothed'' values of the density and the longitudinal and transverse temperatures,
$\{ \ \rho(x),T_{xx}(x),T_{yy}(x) \ \}$. To reduce fluctuations for {\bf Figure 10} we use data for  $8192 = 128 \times 64$ rather than
2048 particles.  These data are smoothed with a properly normalized one-dimensional form of Lucy's short-ranged smooth-particle weight function\cite{b26},
$$
w(x,h) = (5/4hL)(1 - 6z^2 + 8z^3 - 3 z^4) \ ; \ z \equiv (|x|/h) \rightarrow \int_{-\infty}^{+\infty}dx\int_{-L/2}^{+L/2} w(x)dy  \equiv 1 \ .
$$
$L=L_y$ is the height of the system. The weight function vanishes for $|x|>h$. In the initial hot fluid the 8192-particle system
length was $L_x = 128\sqrt{(3/8)}$, reflecting
both the spacing of close-packed triangular-lattice rows and a density twice the close-packed,
$\rho_{\rm initial} = 2\sqrt{(4/3)} = 2.3094$.
The continuum number density at an $x$ grid point $\rho(x_g)$ is given by the integrated density (delta functions) of particles nearby
in their $x$ coordinate, $\rho(x)$:
$$
\rho(x_g) \equiv \sum_i^N w(x_i-x_g) \simeq \int_{-L/2}^{+L/2}dy\int_{x_g-h}^{x_g+h}w(x-x_g)\rho(x)dx \ .  
$$
The smoothing distributes the influence of each particle over a region of width $2h$ in $x$. The kinetic temperatures are  given by
similarly-averaged differences $\langle p^2 \rangle - \langle p \rangle^2$ . {\bf Figure 10} shows these local temperatures as
functions of the local density for a smoothing length h = 3 at the conclusion of the rarefaction simulation. The plot approximates
a straight line from the origin to the point $(\rho,T)=(2.3094,0.115)$. Such a straight line corresponds to an ideal-gas isentrope,
with the product $v\times T$ constant.

\section{Summary and Suggested Research Directions}
 
Lyapunov analyses provide atomistic demonstrations and explanations of the symmetry-breaking instabilities associated
with nonequilibrium states obeying standard classical mechanics.  Developing robust algorithms for stationary shock
and rarefaction waves is a worthy research goal.  We encourage readers to consider these problems. A research goal
stimulated by the present work is to quantify an instability metric.  Such a metric would necessarily depend upon
offset-vector components distinguishing the past from the future. Such a metric should also be related to entropy
production and the Second Law of Thermodynamics.

A Lyapunov analysis of stationary states, as opposed to the transients treated here, is highly desirable.  Steady-state
shockwave simulations, with particles entering at the left and exiting at the right, just as in the stationary view of
{\bf Figure 2}, would make it possible to carry out longtime averages of instability properties. Most likely such an
approach would assign to each particle in a variable-number system private forward and backward vectors, both offset
from the reference trajectory.  These vectors would give pairs of private Lyapunov exponents, $N$ forward and $N$
backward at any time. Histories of these pairs could then be averaged to minimize fluctuations.

The continuum entropy production, depending as it does on gradients of thermodynamic properties, cannot distinguish between
the two time directions.  On the other hand the difference between the instability metrics forward and backward in time,
because they depend only on their ``pasts'', offers the chance better to quantify the relative stability of motions
obeying and disobeying macroscopic thermodynamics.

\section{Acknowledgments}
We very much enjoyed the chance to help honor Berni at his 90th Birthday Symposium at the Livermore Laboratory on 20 August 2015.
In turn, Berni very kindly delivered the keynote address at Bill's 80th Birthday Celebration at Sheffield the following year, on
26 July 2016. Our description of this pedagogical example of the irreversibility inherent in Newtonian dynamics was motivated in
part by email correspondence with Marcus Bannerman and Kris Wojciechowski. We are grateful for their interest.  We are grateful
to the anonymous referee who pointed out some typographical errors in an earlier version of the manuscript and suggested that 
the direction of increasing time be indicated by arrows in Figures 4, 6, and 8.

\pagebreak

\pagebreak

\begin{figure}
\includegraphics[width=1.4 in,angle=+90.]{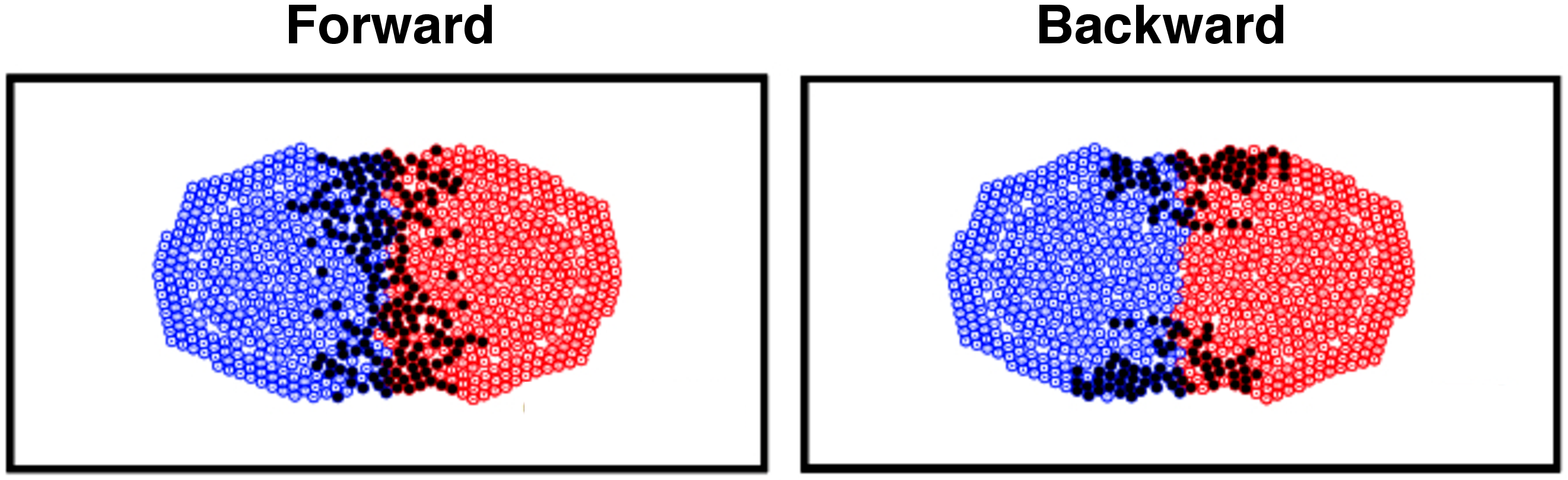}
\caption{
Two identical snapshots from a ``bit-reversible'' precisely-time-reversible Newtonian collision of two solid
400-particle balls\cite{b10}.  The important particles forward and backward in time show that local mechanical instability,
not phase volume, is the mechanism for Second Law irreversibility.
}
\end{figure}

\begin{figure}
\includegraphics[width=2 in,angle=+90.]{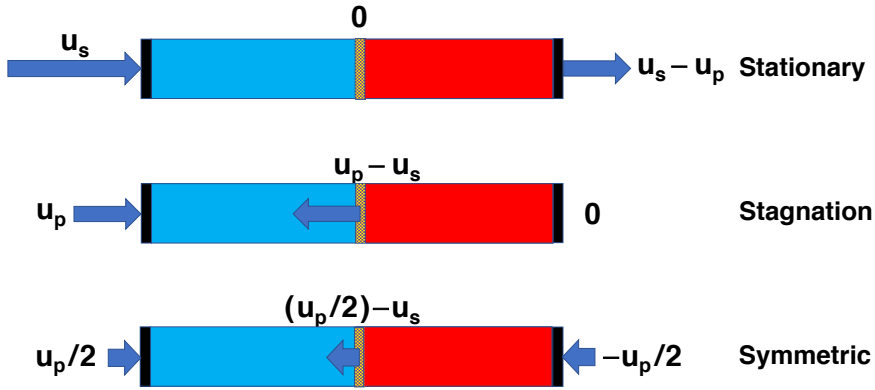}
\caption{
Three mechanisms for generating one-dimensional shockwaves. We use stagnation geometry here. The symmetric mechanism
leads to the Hugoniot Relation $\Delta e = \langle P \rangle \Delta v$, where $\langle P \rangle$ is the
average of the cold and hot pressures and $\Delta v$ is the difference of the cold and hot volumes.
}
\end{figure}

\begin{figure}
\includegraphics[width=1.5 in,angle=90.]{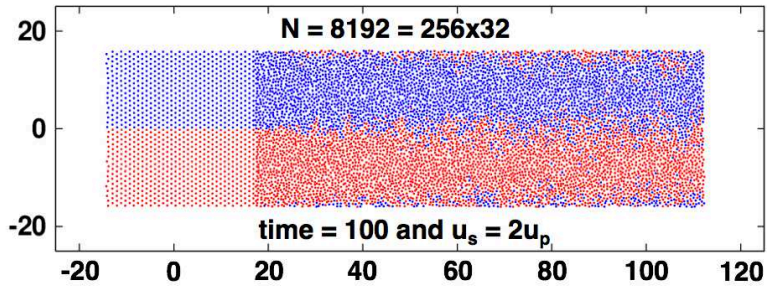}
\caption{
A  one-dimensional leftmoving shockwave. Initially cold solid at density $\sqrt{(4/3)}$ moves rightward at $u_p = 0.97$,
stagnates at a fixed barrier at $x=128\sqrt{(3/4)} = 110.85$, launches a twofold-compressed shockwave leftward, at
$u_p-u_s = -0.97$. Colors show original $y$ values. The timesteps in all of these simulations are equal to 0.01.
}
\end{figure}

\pagebreak

\begin{figure}
\includegraphics[width=3 in,angle=-90.]{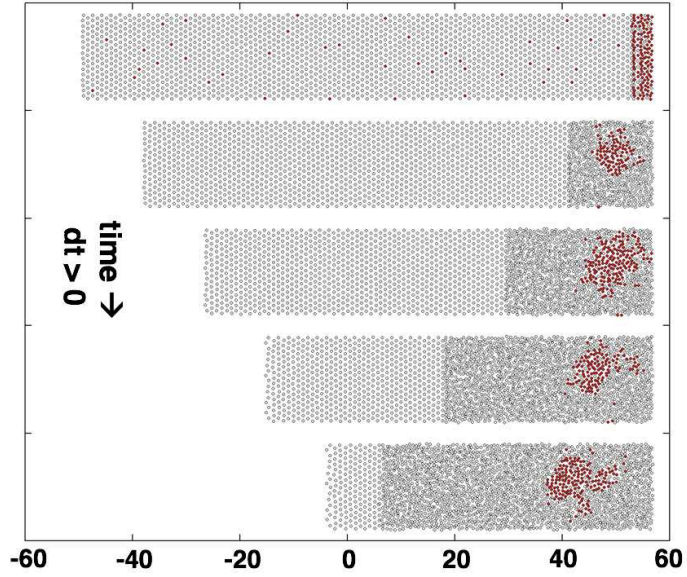}
\caption{
Runge-Kutta shockwave forward in time. The original cold zero-energy zero-temperature specimen, moving rightward at speed 0.97,
had a length of $128\sqrt{3/4}$ and a height of 16.  There are 2048 particles with an initial nearest-neighbor spacing of unity.
The snapshots taken forward in time correspond to times of 6, 18, 30, 42, and 54.  The motion is reversed at time 60.}
\end{figure}

\begin{figure}
\includegraphics[width=3 in,angle=-90.]{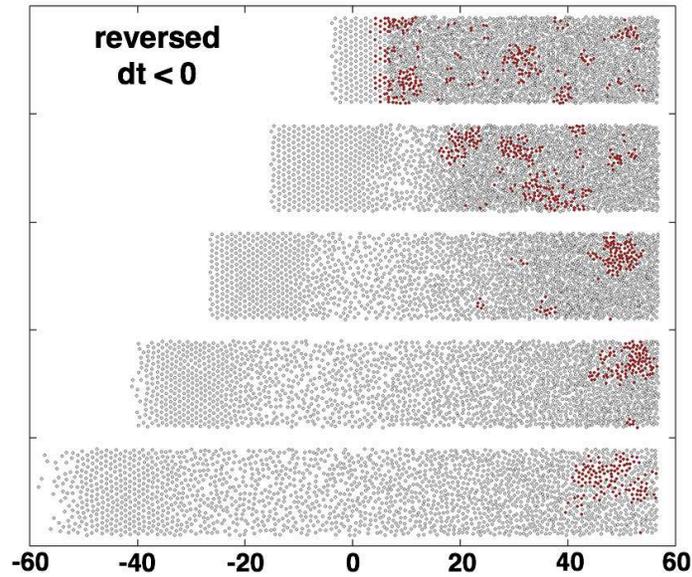}
\caption{
Reversed Runge-Kutta shockwave breaks up and yields a rarefaction wave.  Evidently the reversed shockwave structure is highly
unstable. The times here correspond to those in Figure 4.
}
\end{figure}

\pagebreak

\begin{figure}
\includegraphics[width=3 in,angle=-90.]{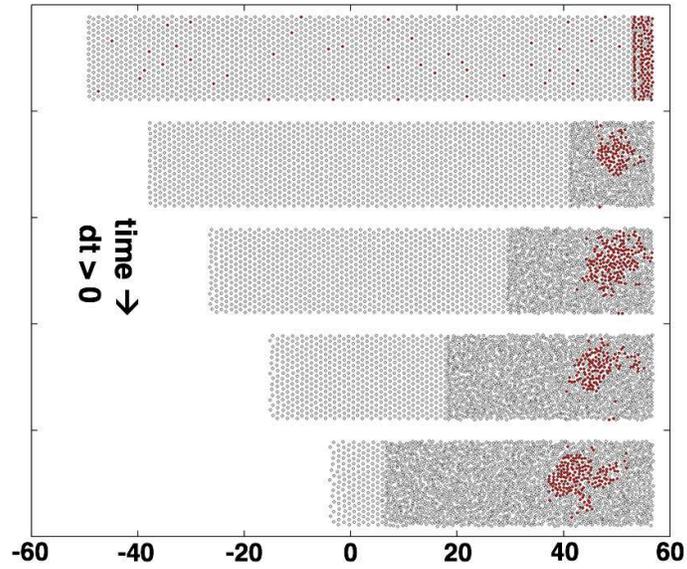}
\caption{
A precisely reversible shockwave stored during propagation forward in time. Here the coordinates and momenta are stored, matching Figure
4. The reversed shockwave structure has been stored for use in Figure 7.   
}
\end{figure}

\begin{figure}
\includegraphics[width=3 in,angle=-90.]{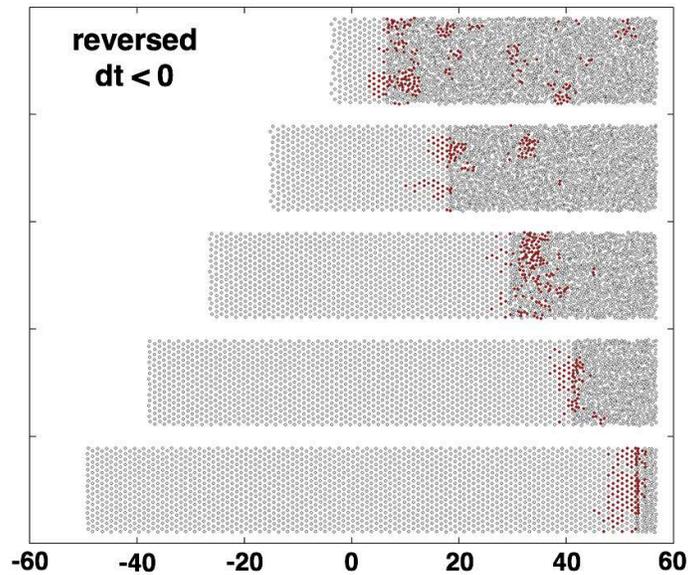}
\caption{
The stored trajectory of Figure 6 is played backward, with the velocities reversed.  Notice that the important particles, colored
brown, are concentrated near the reversed shockwave, indicating its enhanced instability. After reversal at time 60 time decreases
through the snapshot times of 54, 42, 30, 18, and 6.
}
\end{figure}

\pagebreak

\begin{figure}
\includegraphics[width=4 in,angle=-0.]{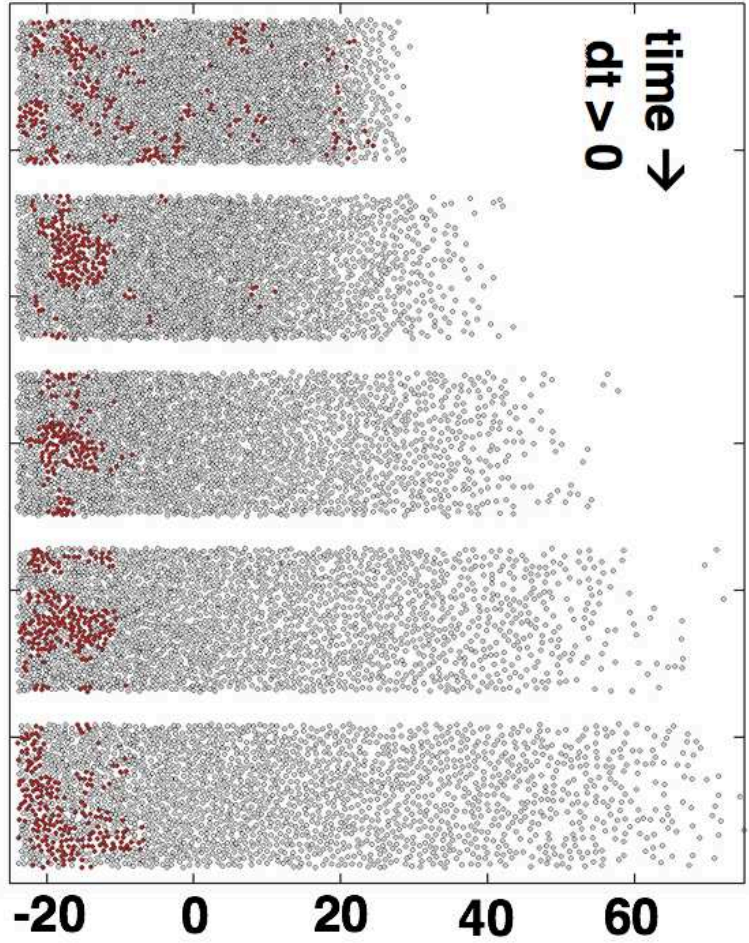}
\caption{
Here and in Figure 9 snapshots at times of 4, 12, 20, 28, and 36 show that the important particles accumulate in clumps near the
left wall, which feels the recoil pressure reacting to the rarefaction fan's motion to the right.  Here the original length of the
2048-particle hot specimen, with temperature 0.115 and density $2\sqrt{(4/3)}$, was $32\sqrt{(2)}$ with height $8\sqrt{(6)}$.
Unlike the shockwave problem the important particles in both time directions occur near the warmer lefthand boundary.
}
\end{figure}

\begin{figure}
\includegraphics[width=4 in,angle=-0.]{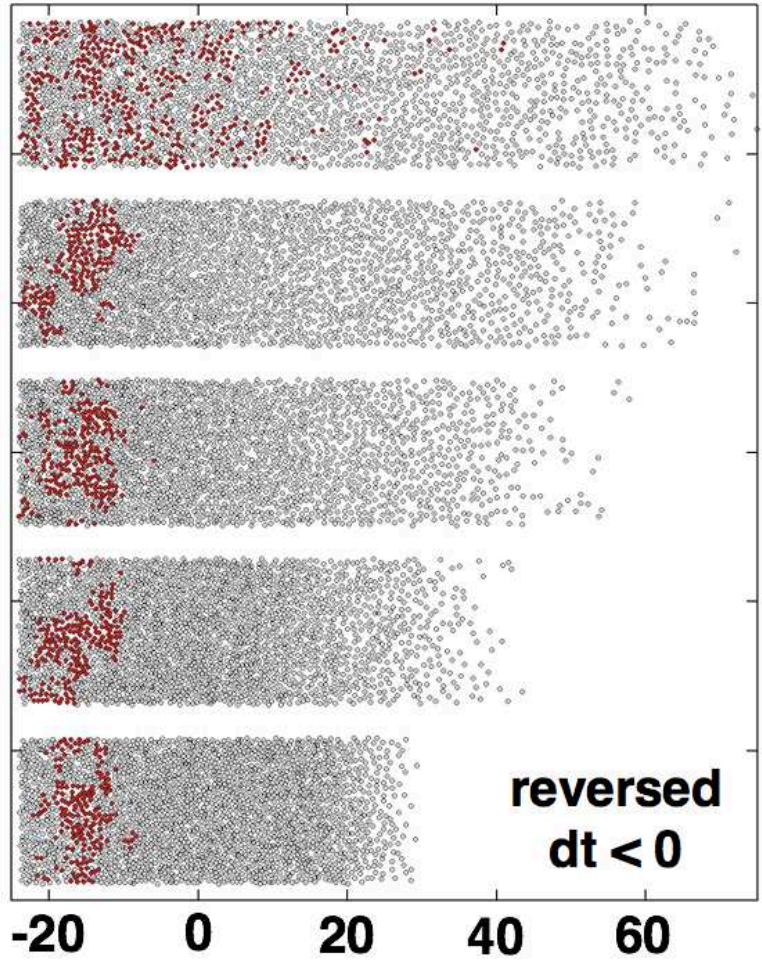}
\caption{
The stored precisely-reversed rarefaction wave of Figure 8 analyzed backward in time. As before the important particles are colored brown.
}
\end{figure}

\pagebreak
\begin{figure}
\includegraphics[width=4
 in,angle=-90.]{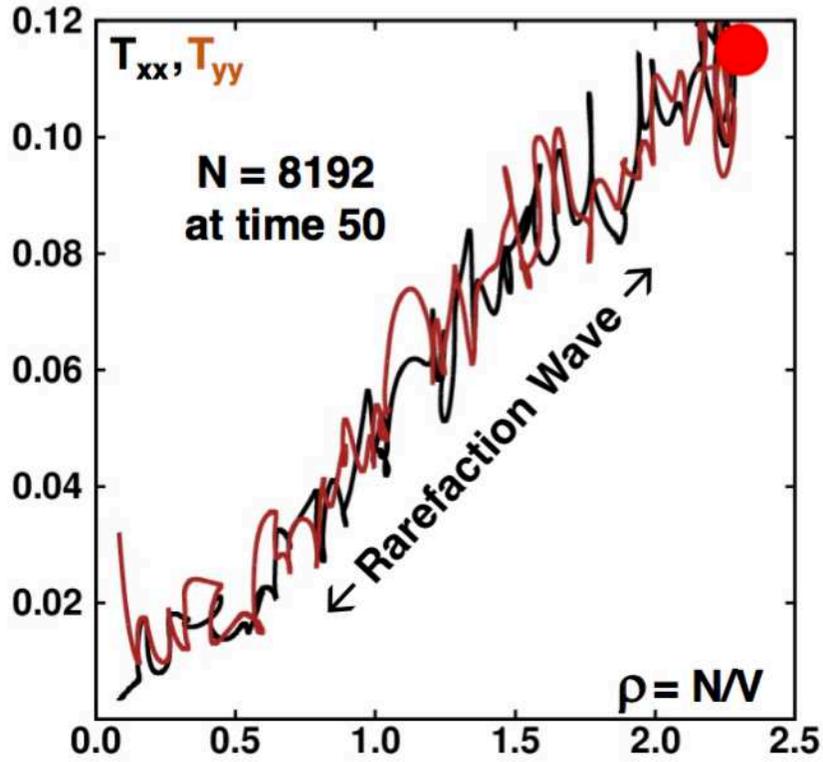}
\caption{
Temperatures and Density in a Rarefaction Wave.  To reduce fluctuations 8192 particles were used. The similarity of the
longitudinal and transverse temperatures is remarkable.  The red dot at the upper right indicates the initial thermodynamic
state imposed by Nos\'e-Hoover dynamics.
}
\end{figure}

\end{document}